\documentclass[aps,prb,reprint,twocolumn,showpacs,floatfix,superscriptaddress,nofootinbib]{revtex4}
\usepackage{graphicx}
\usepackage{amssymb}
\usepackage{amsmath}
\usepackage{lmodern}
\usepackage{color}
\usepackage{hyperref}
\usepackage{empheq}
\usepackage{epstopdf}
\begin{document}

\title{Symmetry of electron  bands in graphene: (nearly) free electron  vs. tight-binding}

\author{Eugene Kogan}
\email{Eugene.Kogan@biu.ac.il}
\affiliation{Jack and Pearl Resnick Institute, Department of Physics, Bar-Ilan University, Ramat-Gan 52900, Israel}
\affiliation{Max-Planck-Institut f\"{u}r Physik komplexer Systeme,  Dresden 01187, Germany}

\author{Vyacheslav M. Silkin}
\email{vyacheslav.silkin@ehu.es}
%\affiliation{Max-Planck-Institut fur Physik komplexer Systeme,  Dresden 01187, Germany}
\affiliation{Donostia International Physics Center (DIPC), Paseo de Manuel Lardizabal 4, 20018 San Sebasti\'an/Donostia, Spain}
\affiliation{Departamento de Pol\'{\i}meros y Materiales Avanzados: F\'{\i}sica, Qu\'{\i}mica y Tecnolog\'{\i}a,
 Facultad de Ciencias Qu\'{\i}micas, Universidad del Pa\'{\i}s Vasco (UPV/EHU),
Apartado 1072, 20080 San Sebasti\'an/Donostia, Basque Country, Spain}
\affiliation{IKERBASQUE, Basque Foundation for Science, 48009 Bilbao, Basque Country, Spain}

\date{\today}

\begin{abstract}
We present the symmetry labelling of all electron bands  in graphene obtained by combining
numerical band calculations and analytical analysis based on group theory. The  latter
was performed both  in the framework of the (nearly) free electron model, or  in  the framework of the  tight-binding model. The predictions about relative positions of the bands which can be made on the basis of each of the models just using the group theory (and additional simple qualitative arguments, if necessary) are complimentary.
\end{abstract}

%\pacs{75.50.Mm, 72.15.Qm, 03.75.Mn}

\maketitle

\section{Introduction}

The electronic band structure of graphite and a graphite
monolayer, called graphene, was a subject of intense study since analytic calculation of Wallace employing a tight-binding model (TBM).\cite{wapr47}
In particular, understanding of the symmetries
of the electronic energy bands in graphene was of crucial
importance. First symmetry classification of them in graphene
was presented  by Lomer in his seminal paper \cite{1}. Later on the subject was developed in numerous publications.\cite{2,3,4,5,6}
Despite of this, band theory and group-theoretical analysis of two-dimensional hexagonal materials in general, and graphene in particular,
continues to attract attention in the very recent years.\cite{gobaacsn10,vodeprl12,chowdhury,ayria,campo,minami,adorno,pisarra,schoop,cano,islam,
kruthoff,bouhon,ferreira}.
% by Slonczewski and
%Weiss, Dresselhaus and Dresselhaus \cite{3},  Bassani and
%Parravicini \cite{4}, Malard et al \cite{5}, Manes \cite{6} and others.

\begin{figure}[h]
\includegraphics[width=1\columnwidth]{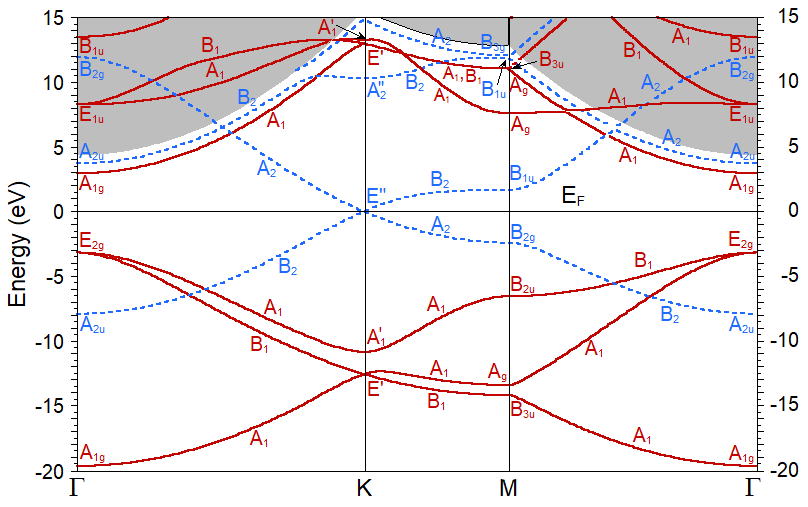}
\caption{
Calculated graphene band structure with labeling of states at the symmetry points and symmetry directions of the Brillouin zone. The $\sigma$-like bands  are plotted with red solid lines and  the  $\pi$-like
bands - with blue dashed lines.
%Occupied bands are plotted with solid lines, unoccupied -
%with dashed lines.
The grey area corresponds to the vacuum continuum states. The black horizontal line shows the Fermi
energy position.
%$[$The bands plotted there are in fact  scattering resonances of graphene
%\cite{Nazarov-13,Wicki-16}. ?????$]$
}
\label{fig:bands}
\end{figure}

In our studies of electronic bands in graphene
we combined numerical band calculations with the analytical symmetry analysis of the bands.\cite{nazarov,ks,sk}
In Fig. \ref{fig:bands} we reproduce
 the results of the band structure calculations with symmetry labelling following our previous papers,\cite{ks,sk} where the details can be found.
Here we just remind about the distinction between   the $\sigma$ and the $\pi$ bands (the former being even with respect to reflection in the plane of graphene,
the latter - odd.) Attention is traditionally attracted to the $\pi$ bands merging at the Fermi level.\cite{neto}
However, we are interested in all the bands (and even in the lowest scattering resonances).

By the analytical symmetry analysis we mean reproduction of the electron band symmetry
without solving any differential or even algebraic equations, but just using the group theory (and additional simple qualitative arguments, if necessary).
To make such analysis possible, one must chose  a model as simple as possible.

There are two alternative approaches  to the analysis, both presented, for example, in the book by Kittel.\cite{kittel}
One can use either  TBM, or the (nearly) free electron model (FEM); note that in spite being just opposite to each other, as a rule, these two approaches give the same result for the  bands symmetry.\cite{kittel,sutton}

The minimalistic tight-binding model,  with four orbitals on each atoms ($|2s>,|2p>$), correctly describes the symmetry of
all the occupied bands and the unoccupied $\pi$ band touching the Fermi level.
We  call these bands the TBM bands.

However, the minimalistic TBM  doesn't describe correctly the other unoccupied bands. The latter are also differ from the TBM bands in their  dispersion law and  localization with respect to the graphene plane.
This is why we called them  the FEM bands.\cite{ks,sk}
To  describe the symmetry of all the bands, we used a hybrid approach,
combining TBM and FEM.\cite{sk} \textcolor[rgb]{0.50,0.00,0.00} {For a very recent review see Ref. \onlinecite{naumis}.}

In the present paper we want now to draw attention to the fact, that symmetry analysis
of all the bands can be performed alternatively within each of the models - either FEM or TBM
(for the latter at the price of extending the basis of atomic orbitals). We compare the predictions (and the predictive power) of the models.

To understand the symmetry classification of the bands, one should remember that the group of wave vector ${\bf k}$  at the $\Gamma$ point is $D_{6h}$;
  at the $K$ point -- $D_{3h}$;  at the point $M$ -- $D_{2h}$.
The group of wave vector ${\bf k}$  at each of the lines constituting triangle $\Gamma-K-M$  is $C_{2v}$.\cite{thomsen,dresselhaus}
Representations of the groups can be found in the book by Landau and  Lifshitz.\cite{landau}
One of rotations $U_2$ for the $D_{6h}$ group is about the  direction $\Gamma - K$.  Rotation  $C_2^z$  for the $D_{2h}$ group is about the normal to graphene plane, rotation
 $C_2^x$  - about the $\Gamma - M$ line.
Reflection $\sigma_v$ for the $C_{2v}$ groups is relative to the plane of graphene.

\section{(Nearly) free electron model}
\label{ki}

For the sake of the symmetry analysis, we present the wave functions  of all the bands in the factorised form
\begin{eqnarray}
\label{fm}
\psi_{\bf k}(x,y,z)=f_{\bf k}(z)\phi_{\bf k}(x,y),
\end{eqnarray}
where $\phi_{\bf k}(x,y)$ are linear combinations of appropriate plane waves,  and the functions $f(z)$ are determined by the boundary conditions
$\lim_{z\to\pm\infty}f_{\bf k}(z))=0$.
For the $\sigma$ band $f(z)$ is an even function, and for   the  $\pi$ band -- an odd one.
Analysis of the representations of the groups realised by the plane waves is presented in our previous publications.\cite{ks,sk}
Notice that the model can equally well incorporate both the TBM bands, localized in graphene, and the FEM bands, having long vacuum tails. The distinction between the two kinds of bands will be reflected in difference between the corresponding functions $f(z)$.

The model potential which would correspond to our choice of the wave functions for the FEM  is the sum of two potentials:
$x,y$ independent and $z$-dependent strong potential $V_1(z)$ which localizes the electron states near  the graphene plane,
and   weak potentials  $V_2(x,y,z)$ which have graphene lattice symmetry in the $x,y$ plane.
Probably, to take into account the existence of carbon ion cores, it would be more correct to consider $\psi$ (and $\phi$) as a kind of orthogonalized plane wave, and $V_2(x,y,z)$ as  pseudo potentials.
If we compare the two lowest $\sigma$ bands with the two lowest $\pi$ ones on Fig. \ref{fig:bands}, the idea to treat the same way both classes of bands
looks quite natural.

\begin{figure}[h]
\includegraphics[width= .95\columnwidth]{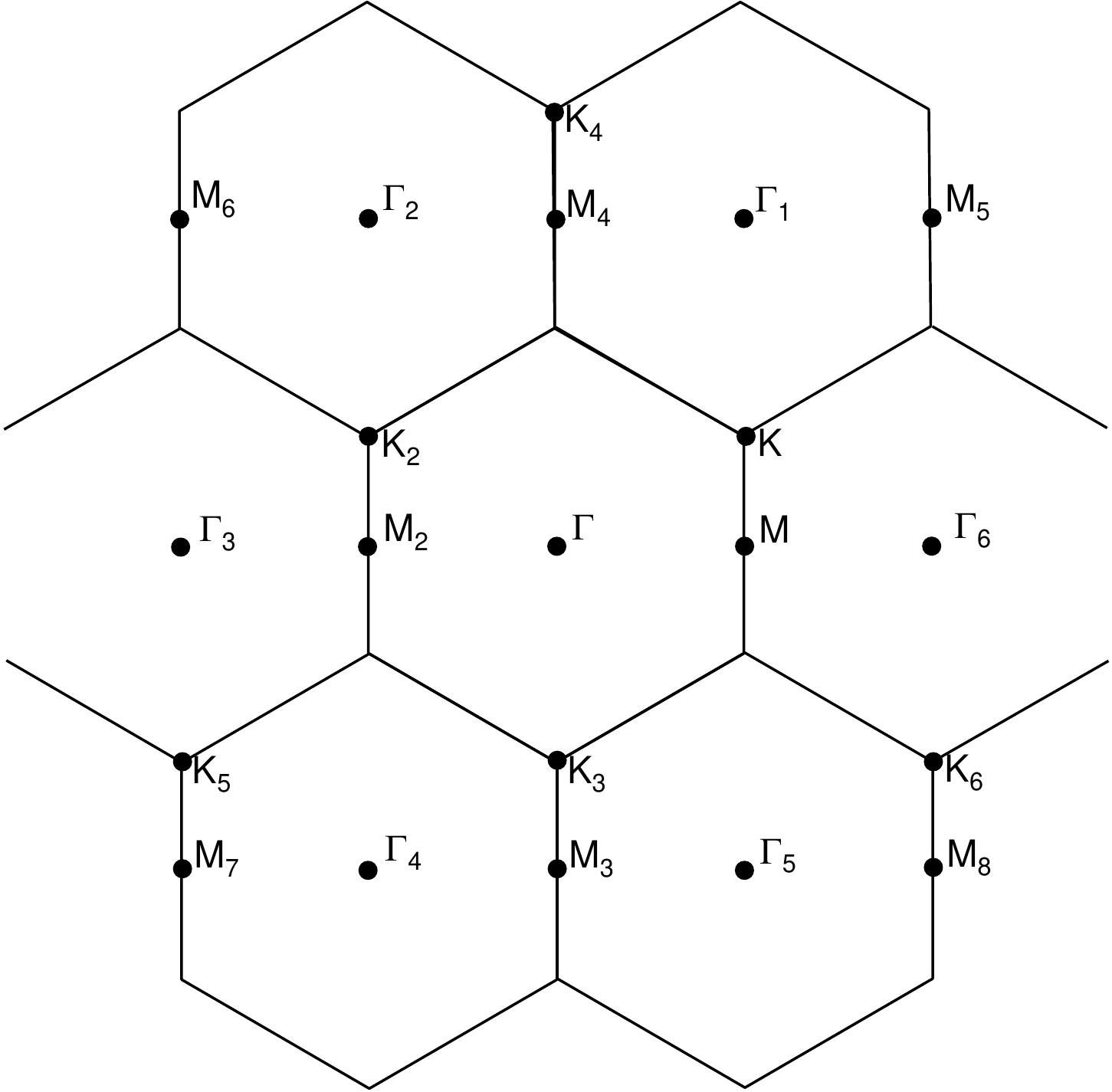}
\caption{\label{fig:bandsn}  Extended reciprocal lattice for the honeycomb lattice. The meaning of the symbols see in the text. }
\end{figure}

Extended reciprocal lattice for the honeycomb lattice, we will use, is presented on Fig. \ref{fig:bandsn}.
Wave functions $\phi$ of the lowest energy states inside the Brillouin zone (BZ) are just plane waves, at the boundaries of the zone -  combinations
of two  plane waves, and at the band vertexes - of three plane waves.\cite{kittel}
Weak lattice potential should lead to small splitting within a doublet or triplet.
Fig. \ref{fig:bands}, with its the lowest singlet at the $\Gamma$ point, lowest doublet at the $K-M$ line and lowest triplet at the $K$ point,  certainly speaks in favor of the applicability of the approach.

More specifically, a single point (or line) in the reduced scheme corresponds to infinite number of points (or lines) in the extended scheme.
Thus to a single point $\Gamma$ in the reduced scheme, in the extended scheme there correspond the $\{\Gamma$ points, $\Gamma_{1},\dots,\Gamma_6\}$, etc.,
to the $K$ point - the points $\{K,K_2,K_3\}$,   $\{K_4,K_5,K_6\}$, etc., to the $M$ point - the points $\{M,M_2\}$,   $\{M_3,M_4\}$, $\{M_5,M_6\}$, etc.
To a single line $K_M$ in the reduced scheme, in the extended scheme there correspond the lines
$\{K-M,K_2-M_\}$, $\{K_3-M_3,K_4-M_4\}$, $\{K_5-M_7,K_6-M_8\}$ etc.

At the $\Gamma$ point, the plane wave $\phi_{\bf \Gamma}(x,y)$ equal identically to 1  realizes representation $A_{1g}$ for even $f(z)$ and representation $A_{2u}$ for odd $f(z)$.
At the line  $\Gamma - K$  the only basis plane wave $\phi_{\bf k}(x,y)$
realizes representation $A_{1}$ for even $f(z)$ and representation $B_{1}$ for odd $f(z)$.

At the $K$ point, $\phi_{\bf K}(x,y)$ is a linear combination of three plane waves
$e^{i{\bf K}\cdot {\bf r}}$
with the  wave vectors  corresponding to the three equivalent vertices of the hexagon
${\bf K}=\left(\frac{2\pi}{3a},\frac{2\pi}{3\sqrt{3}a}\right)$, ${\bf K}_2=\left(-\frac{2\pi}{3a},\frac{2\pi}{3\sqrt{3}a}\right)$,
${\bf K}_3=\left(0,-\frac{4\pi}{3\sqrt{3}a}\right)$.
For even $f(z)$  the functions $f(z)\times\left\{e^{i{\bf K}_{1}\cdot {\bf r}},e^{i{\bf K}_{2}\cdot {\bf r}},e^{i{\bf K}_{3}\cdot {\bf r}}\right\}$
realize
$A_1'+E'$ representations of $D_{3h}$.
For odd $f(z)$ the functions realize
$A_2''+E''$ representations.
The $\pi$ triplet  can be substantially  higher than the $\sigma$ triplet corresponding to the same plane waves due to the difference  between the energy of  odd and even states in the strong $V(z)$ potential.

To find explicitly splitting of the bands at the $K$ point we should solve
the secular equation, which, taking into account the symmetry and shifting energy by the diagonal
matrix element of the potential, we may write down  in the form
\begin{eqnarray}
\left|\begin{array}{ccc} -E & V_2 & V^* \\ V^* & -E & V \\ V & V^* & -E\end{array}\right|=0,
\end{eqnarray}
where $V_2$ is the matrix element of the potential $V_2(x,y,z)$ between some pair of different states from ${\bf K},{\bf K}_2,{\bf K}_3$.
Using Cardano's formula, we may write down the roots as
\begin{eqnarray}
E=2\text{Re}\left(\sqrt[3]{V^3}\right),
\end{eqnarray}
where Re means real part. From the fact that one of the roots should be doubly degenerate, we come to the conclusion that $V$ is real. (Of course, this  can be obtained in a more direct way, on the basis of potential's symmetry.)
Anyhow, the roots are $E_{1,2}=-V$, $E_3=2V$.
We understand that relative positions of the singlet and the doublet depend upon sign of $V$.
If we assume that the matrix element $V_2$
is  positive, we obtain that in each triplet the doublet should be lower than the singlet.

At the $M$ point, $\phi_{\bf K}(x,y)$ is a linear combination of two plane waves
with the  wave vectors
${\bf M}=\left(\pm\frac{2\pi}{3a},0\right)$.
For even $f(z)$ the  function
$f(z)\times$ (sum or difference of the exponents) realizes $A_g$ and
 $B_{3u}$ representations of $D_{2h}$ respectively.
For odd $f(z)$ the  function
$f(z)\times$ (sum or difference  of the exponents) realizes
 $B_{1u}$ and $B_{2g}$ representation of $D_{2h}$ respectively.

The wave functions of the bands at the $K$ point, higher than the  lowest  triplets, are combinations of the plane waves  with the  wave vectors
${\bf K}_4=\left(0,\frac{8\pi}{3\sqrt{3}a}\right)$, ${\bf K}_5=\left(-\frac{4\pi}{3a},-\frac{4\pi}{3\sqrt{3}a}\right)$,
${\bf K}_6=\left(\frac{4\pi}{3a},-\frac{4\pi}{3\sqrt{3}a}\right)$,
which realize representations identical to those realised by the plane waves with the wave vectors ${\bf K},{\bf K}_2,{\bf K}_3$.
This  explains the second copy of the
representations $A_1'+E'$ we observe at the $K$ point on Fig. \ref{fig:bands}.

The wave functions of the  bands at the $M$ point, higher than the lowest ones,
are combinations of two plane waves  with the  wave vectors
${\bf M}_{3,4}=\left(0,\pm\frac{2\pi}{\sqrt{3}a}\right)$.
For even $f(z)$ the  function
$f(z)\times$ (sum or difference of the exponents) realizes $A_g$ and  $B_{2u}$ representation of $D_{2h}$ respectively.
The bands with these symmetries we see on Fig. \ref{fig:bands}.
%Again we must notice that the large splitting   in the  doublet
%doesn't agree very well with our assumption of weak $V(x,y,z)$ potential.
For odd $f(z)$ the  function
$f(z)\times$ (sum or difference of the exponents) realizes
 $B_{1u}$ and $B_{3g}$ representation of $D_{2h}$ respectively.

To describe still higher bands at the $M$ point we consider  four additional plane waves $e^{i{\bf M}\cdot {\bf r}}$, with the  wave vectors
${\bf M}_{5,\dots,8}=\left(\pm\frac{4\pi}{3a},\pm\frac{2\pi}{\sqrt{3}a}\right)$. These four plane waves, multiplied by even function $f(z)$ realize
$A_g+B_{1g}+B_{2u}+B_{3u}$ representation of the group $D_{2h}$.

The wave functions of the bands at the $\Gamma$ point, higher than the two lowest ones,
correspond to combinations of
6  plane waves  with the  wave vectors $\Gamma_1,\dots,\Gamma_6$,
presented on Fig. {fig:bandsn}, and corresponding to
$\left(\pm\frac{2\pi}{3a},\pm\frac{2\pi}{\sqrt{3}a}\right)$,
$\left(\pm\frac{4\pi}{3a},0\right)$.
For even $f(z)$ the  functions (\ref{fm}) realize $A_{1g}+B_{1u}+E_{2g}+E_{1u}$ representations of the group $D_{6h}$.
To find explicitly splitting of the bands at the $\Gamma$ point we should
diagonalize the Hamiltonian, which in the representation $\Gamma_1,\dots,\Gamma_6$
 is a circulant $6\times 6$ matrix
with the matrix elements
%\begin{widetext}
\begin{eqnarray}
\label{circul}
\hat{H}=\left(\begin{array}{cccccc}
0 & V_2^{(a)} & V_2^{(b)}   &  V_2^{(c)}  &  V_2^{(b)}    &  V_2^{(a)} \\
V_2^{(a)} & 0 & V_2^{(a)}  &  V_2^{(b)}  &  V_2^{(c)}   &  V_2^{(b)} \\
V_2^{(b)} & V_2^{(a)} & 0  &  V_2^{(a)}  &  V_2^{(b)}   &  V_2^{(c)} \\
V_2^{(c)}  & V_2^{(b)} & V_2^{(a)}   &  0  &  V_2^{(a)}    &  V_2^{(b)} \\
V_2^{(b)} & V_2^{(c)} & V_2^{(b)}  &  V_2^{(a)}  &  0   &  V_2^{(a)} \\
V_2^{(a)} & V_2^{(b)} & V_2^{(c)}  &  V_2^{(b)}  &  V_2^{(a)}   &  0 \end{array}\right).
\end{eqnarray}
%\end{widetext}
Note that due to the symmetry of the problem,
there are only 3 different matrix element of $V_2(x,y,z)$: $V_{{\bf \Gamma}_1-{\bf \Gamma}_2}\equiv V_2^{(a)}$, $V_{{\bf \Gamma}_1-{\bf \Gamma}_3}\equiv V_2^{(b)}$,
and $V_{{\bf \Gamma}_1-{\bf \Gamma}_4}\equiv V_2^{(c)}$
(we again  shifted energy by the diagonal
matrix element of the potential).
The  eigenvalues of the matrix (\ref{circul}) are
\begin{eqnarray}
E_i=V_2^{(a)}r_i+V_2^{(b)}r_i^2+V_2^{(c)}r_i^{3}+V_2^{(b)}r_i^4+V_2^{(a)}r_i^5,
\end{eqnarray}
where $r_i$ is one of the distinct solutions of $r^6=1$.
After simple algebra we obtain
\begin{eqnarray}
E_1&=&2V_2^{(a)}+2V_2^{(b)}+V_2^{(c)} \nonumber\\
E_{2,3}&=&V_2^{(a)}-V_2^{(b)}-V_2^{(c)} \nonumber\\
E_{4,5}&=&-V_2^{(a)}-V_2^{(b)}+V_2^{(c)} \nonumber\\
E_{6}&=&-2V_2^{(a)}+2V_2^{(b)}-V_2^{(c)}.
\end{eqnarray}
Considering only $\sigma$ bands, we may say that that the eigenfunction corresponding to $E_1$ realizes $A_{1g}$ representation, the eigenfunctions corresponding to $E_{2,3}$ realize $E_{1u}$ representation, the eigenfunctions corresponding to $E_{3,4}$ realize $E_{2g}$ representation, and the eigenfunction corresponding to $E_6$ realizes $B_{1u}$ representation of the group $D_{6h}$.

If we assume that the largest, by absolute value, matrix elements of the potential $V(x,y,z)$ are between the states, with the opposite wave vectors,
and negative, we obtain that  the three lowest bands are even with respect to rotations by an angle $\pi$ about the $z$ axis, perpendicular to the graphene plane, and the three others are odd. That is sextuplet is divided into two triplets: the lower one - $A_{1g}+E_{2g}$ and the higher one - $B_{1u}+E_{1u}$.
%Within each triplet the order of bands is the doublet below and the singlet above, due to %the same reasons it happenes at the point $K$.

%For odd $f(z)$ the  functions $f(z)\times$ plane wave realize $A_{2u}+B_{2g}+E_{1u}+E_{2g}$ %representations of the group $D_{6h}$.
%The band $A_{2u}$ we observe on Fig. \ref{fig:bands}. The other bands are swallowed by the %continuum.

On the line $K -M$ in the reduced scheme, the lowest doublet would corresponds to two plane waves with the wave vectors on the  lines
$K-M$ and $K_2-M_2$
For even $f(z)$, the function $f(z)\times$ (sum of the exponents)
 realizes $A_1$  representations,  and the function $f(z)\times$ (difference of the exponents) realizes $B_1$
representation  of $C_{2v}$.
For odd $f(z)$, the function $f(z)\times$ (sum of the exponents) realizes
$B_2$ representation,  and the function
$f(z)\times$ (difference of the exponents) realizes
and $A_2$ representation of the group.

The third band corresponds to the single plane waves with the wave vectors on the  lines $K_3-M_3$,
and the forth band corresponds to the single plane waves with the wave vectors on the  lines $K_4-M_4$.
Both realize representation $A_1$.

Then comes doublet corresponding to the plane waves with the wave vectors on the line $K_5-M_7$ and $K_6-M_8$.
From the point of symmetry it should be identical to the first doublet.

\section{Tight-binding model}
\label{ti}

In the frame of the tight-binding model we look for the solution of the Schr\"{o}dinger equation as a linear combination of the functions
\begin{eqnarray}
\label{tb}
\psi_{\beta;{\bf k}}^j=\sum_{{\bf R}_j} e^{i{\bf k\cdot R}_j}\psi_{\beta}\left({\bf r}-{\bf R}_j\right),
\end{eqnarray}
where $\psi_\beta$ are atomic orbitals, $j=A,B$ labels the sub-lattices, and  ${\bf R}_j$ is the radius vector of an atom in the sublattice $j$.
(Notice that we assume only symmetry of the basis functions with respect to rotations and reflections; the question how these functions  are
related to the atomic functions of the isolated  atom is irrelevant.)

A  symmetry transformation of the functions $ \psi_{\beta;{\bf k}}^j$ is a direct product of two transformations: the transformation of the sub-lattice functions $\phi^{A,B}_{{\bf k}}$, where
\begin{eqnarray}
\label{2}
\phi_{\bf k}^j=\sum_{{\bf R}_j} e^{i{\bf k\cdot R}_j},
\end{eqnarray}
and the transformation of the orbitals $\psi_{\beta}$. Thus the representations realized by the functions (\ref{tb}) will be the direct product of two representations.  One should pay attention that the wave vector ${\bf k}$ in Eqs. (\ref{tb}) and (\ref{2}) is
reduced to the first BZ, while the wave vector in Eq. (\ref{fm}) was considered as belonging to the infinite plane (extended zone scheme).

We'll start from summing up the results of the symmetry analysis in the framework of the TBM obtained in our previous publications, when
the  basis included only four atomic orbitals: $|s,p>$.\cite{nazarov,ks,sk} The $\sigma$ bands are constructed from the $|2s,2p_{x,y}>$ orbitals,
and the $\pi$ bands are constructed from the $|p_z>$ orbitals.
At the $\Gamma$ point the representations realised by the $\sigma$ bands are
$A_{1g}+B_{1u}$ (constructed from the $|s>$ orbitals) and $E_{1u}+E_{2g}$ (constructed from the $|p_{x,y}>$ orbitals); the representations realised by the $\pi$ bands are $A_{2u}+B_{2g}$.
 At the $K$ point the representations realised by the $\sigma$ bands are
$A_{1}'+A_{2}'$  (constructed from the $|p_{x,y}>$ orbitals) and twice  $E'$ (constructed from the $|s,p_{x,y}>$ orbitals);  the representation realised by the $\pi$ bands is $E''$.
 At the $M$ point the representations realised by the $\sigma$ bands are $A_g+B_{3u}$ (constructed from the $|s>$ orbitals,
same representations  constructed from the $|p_{x}>$ orbitals, and
$B_{1g}+B_{2u}$ (constructed from the $|p_y>$ orbitals); the representations realised by the $\pi$ bands are $B_{1u}+B_{2g}$.

Just by counting the number of bands on Fig. \ref{fig:bands} we realize that the basis of atomic orbitals should be expanded to describe
additional  bands. Actually, the necessity to extend the basis
for accurate description  of  the occupied bands,
comparable to the result of calculations based on plane waves
is well known (traditionally one chooses two sets of
s, p and one set of d) atom-centered basis functions
based on the atomic orbitals.
However,
this choice yields a wrong description of the first
unoccupied bands, which start about 3.25 eV above the
Fermi level and are parabolic around the BZ
center, $\Gamma$.\cite{stewart} These bands have
 long expansion into the vacuum, and are strongly influenced by
the image-potential tail\cite{silkin} with
even and odd mirror symmetry in the graphene plane. Moreover, they can be easily influenced by applied electric field,\cite{bobaprl10,bosinjp10} \textcolor[rgb]{0.50,0.00,0.00} {adsorbate deposition,\cite{weliapl08,wesaprm17,hecapla19} or} transformed upon variation of the graphene sheet shape.\cite{fezhs08}
Notice that  the first two unoccupied states  are
important for e.g. the description of interlayer states,
reactivity, intercalation,\cite{posternak,agapito} and tunneling into
graphene, where the inelastic phonon scattering plays a
dominant role.\cite{zhang,wehling}
To overcome this defect, there was  presented an interesting idea to add long-range orbitals to the minimalistic $|2s,2p>$ basis. \cite{papior}
\textcolor[rgb]{0.50,0.00,0.00} {Notice that the main dynamical effect defining the long tail bands proposed in Ref. \onlinecite{papior} is precisely the image potential experienced by an electron due to its image charge in the graphene sheet.\cite{silkin,nifajpcm14,argunjp15}}

%(being in odd wavefunctions  with respect to the graphene plane), which was   considered in reference \cite{silkin}.   See also  Ref.  \cite{hecapla19},   in which the image potential  defining the free bands is argued to allow the possibility of graphene to become metallic under water doping.

Our paper is mostly devoted
to FEM, and in the spirit of our emphasis of simple models, we decided, while considering the TBM for comparison, just to add to the minimalistic basis atomic orbitals one by one, to understand
which minimal additions are necessary to describe the symmetry of all calculated bands.
Analysis of the TBM with the above mentioned long-range orbitals will be the subject of our next publication.

The first choice is obvious -
$|3s>$ atomic orbital, to describe redundant $\sigma$ band, and $|3p_z>$ atomic orbital, to describe redundant $\pi$ band.
As far as the symmetry  is concerned,  the orbitals give  copies of the bands constructed from $|2s>$ and $|2p_z>$ atomic orbitals.
The fact, that the symmetry of the two lowest unoccupied bands at the $\Gamma$ point is identical to the
symmetry of the two lowest occupied ones, speaks in favor of such choice.

However, there is
a problem with the unoccupied $\pi$ band at the $K$ point. The $|3p_{z}>$ atomic orbitals, like $|2p_{z}>$ orbitals, give doubly
degenerate band at the point. To solve the problem we have to introduce
 $|3d>$ orbitals. In fact,
expanding the $D^{(2)}$ representation of the rotations group, the orbitals realise, with respect to irreducible representations of the group $D_{3h}$,\cite{landau}  we obtain
\begin{eqnarray}
D^{(2)}=A_{1}'+E'+E''.
\end{eqnarray}
We can chose the bases of the representations respectively as
\begin{eqnarray}
\left|d_{z^2}\right>;\;\;\left|d_{x^2-y^2}\right>,\left|d_{xy}\right>;\;\;\left|d_{xz}\right>,\left|d_{yz}\right>.
\end{eqnarray}
The $\pi$ band should be constructed from the last two orbitals.
The functions $\phi_{\bf K}^{A,B}$ realize $E'$ representation of the group $D_{3h}$ .
Thus $\pi$ bands at the $K$ point  realise the following representations:
\begin{eqnarray}
\label{g02b}
E'\times E''&=&A_1''+A_2''+E''.
\end{eqnarray}
Thus  the calculated $A_2''$  band is accounted for.

\section{Comparison between the band calculations and the predictions of the models}

\begin{figure}[h]
%\centering
\includegraphics[width= 1\columnwidth]{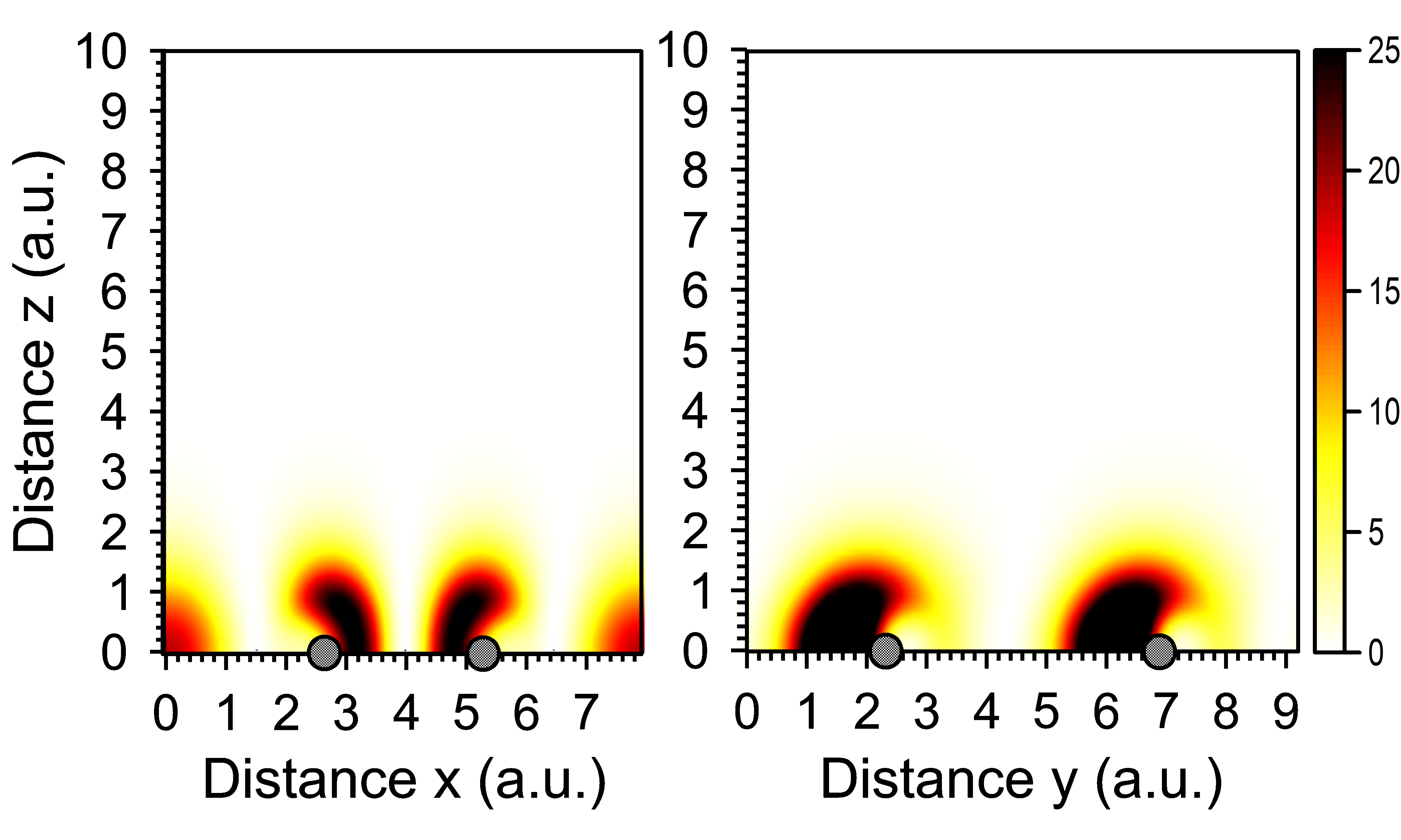}
\caption{\label{Charge06}   Charge-density distribution (in arbitrary units) in $x=0$ plane for the sixth band at the $M$ point. Filled dots show the carbon ion positions. }
\end{figure}

\begin{figure}[h]
%\centering
\hskip -.5cm
  \begin{minipage}[b]{0.25\textwidth}
  \includegraphics[width= \textwidth]{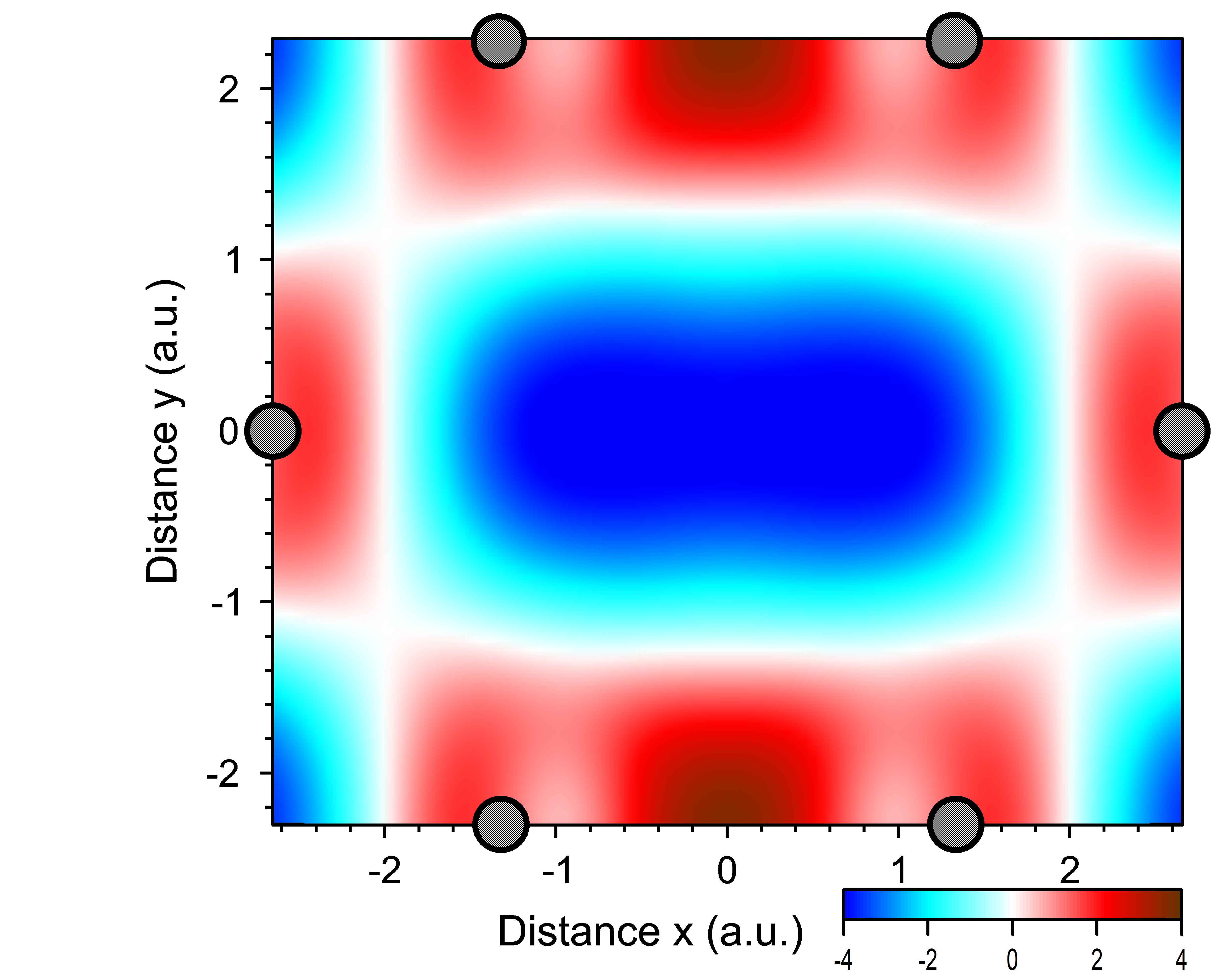}
\end{minipage}
  \hfill
  \begin{minipage}[b]{0.25\textwidth}
\includegraphics[width= \textwidth]{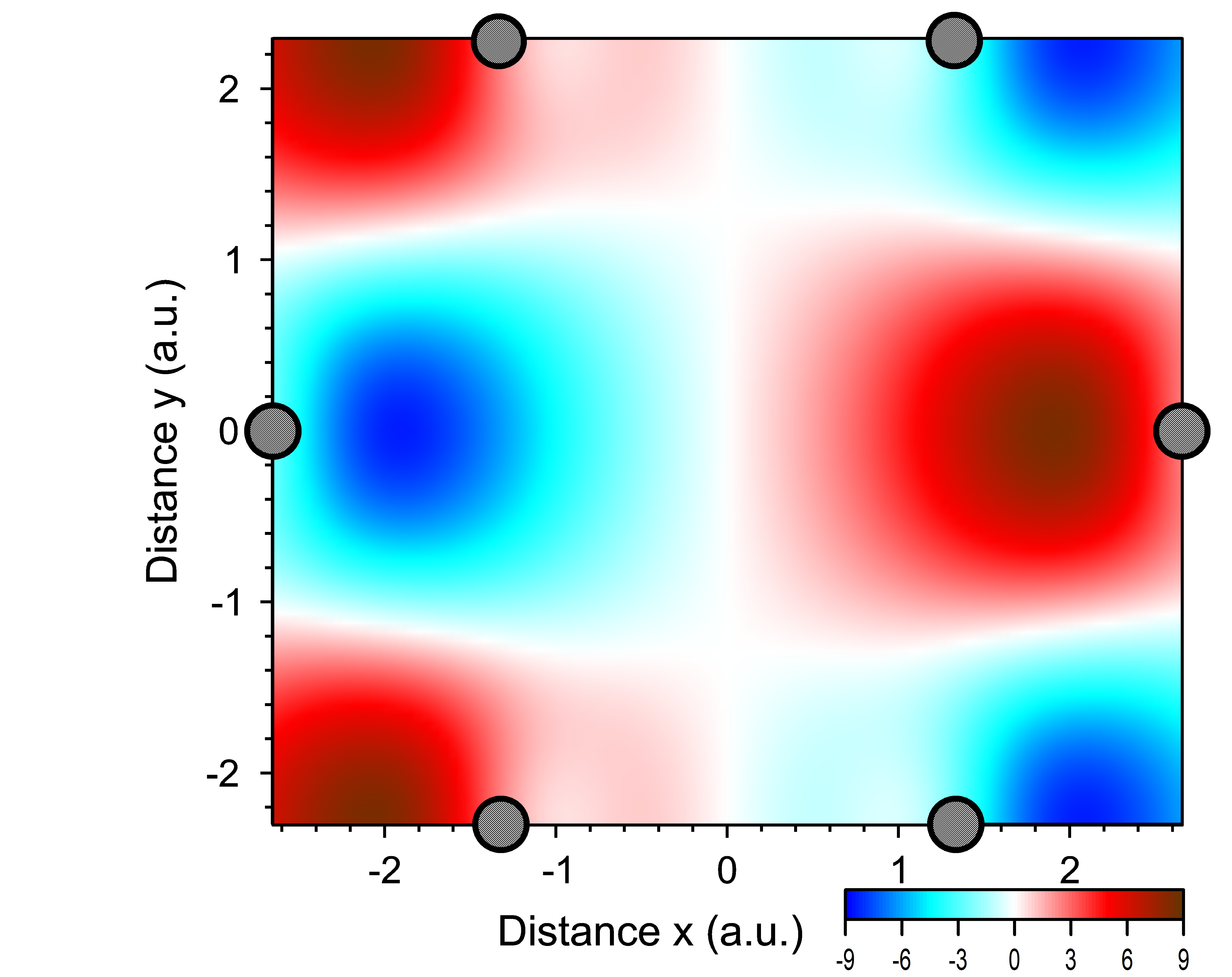}
\end{minipage}
\caption{\label{Psi06}   Real (left) and imaginary (right) parts of the wave function (in arbitrary
units) at the $z = 0$  plane for the sixth band at the $M$ point. }
\end{figure}

\begin{figure}[h]
%\centering
\includegraphics[width= 1\columnwidth]{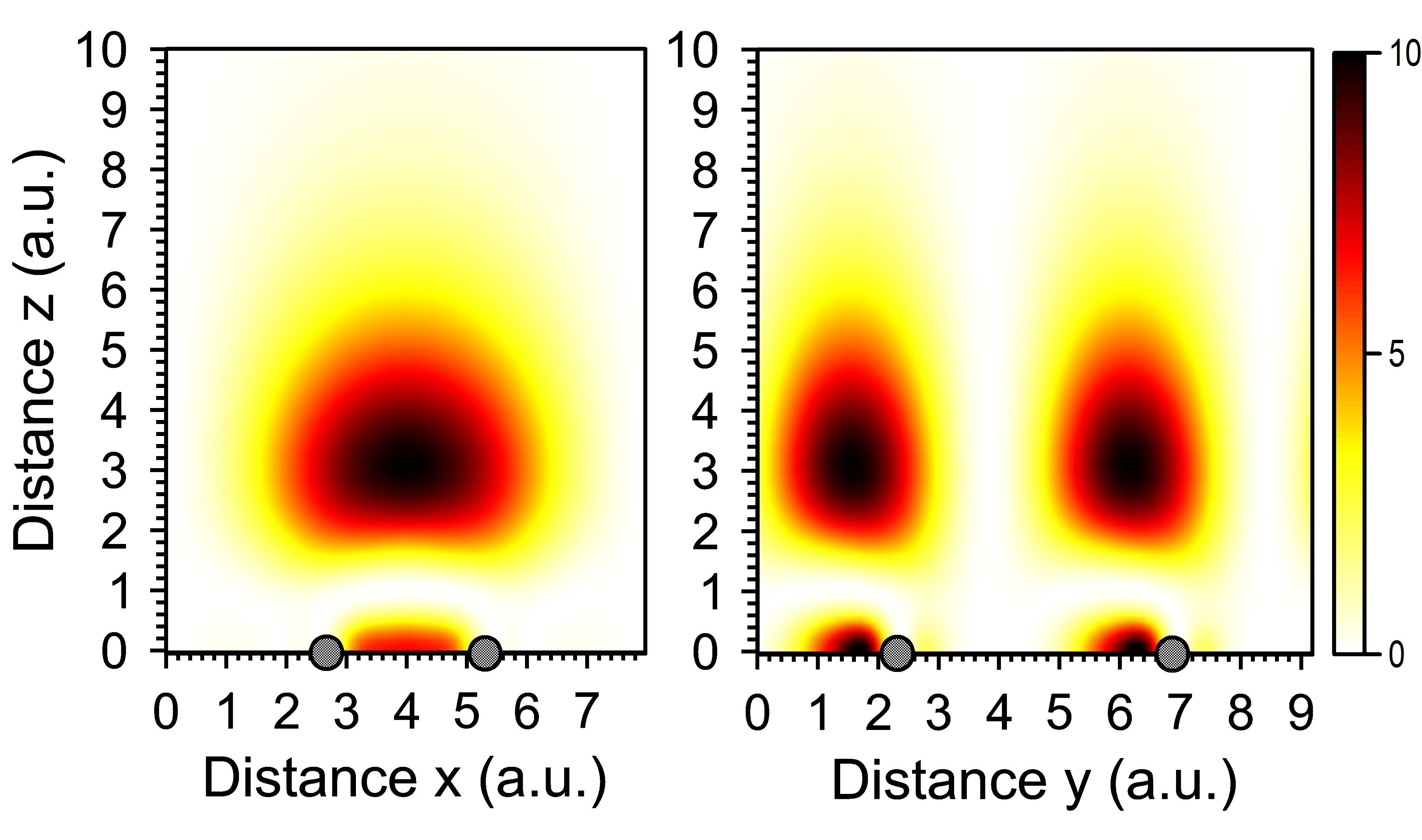}
\caption{\label{Charge07}   Charge-density distribution (in arbitrary units) in $x=0$ plane for the seventh band at the $M$ point. Filled dots show the carbon ion positions. }
\end{figure}

\begin{figure}[h]
%\centering
\hskip -.5cm
  \begin{minipage}[b]{0.25\textwidth}
  \includegraphics[width= \textwidth]{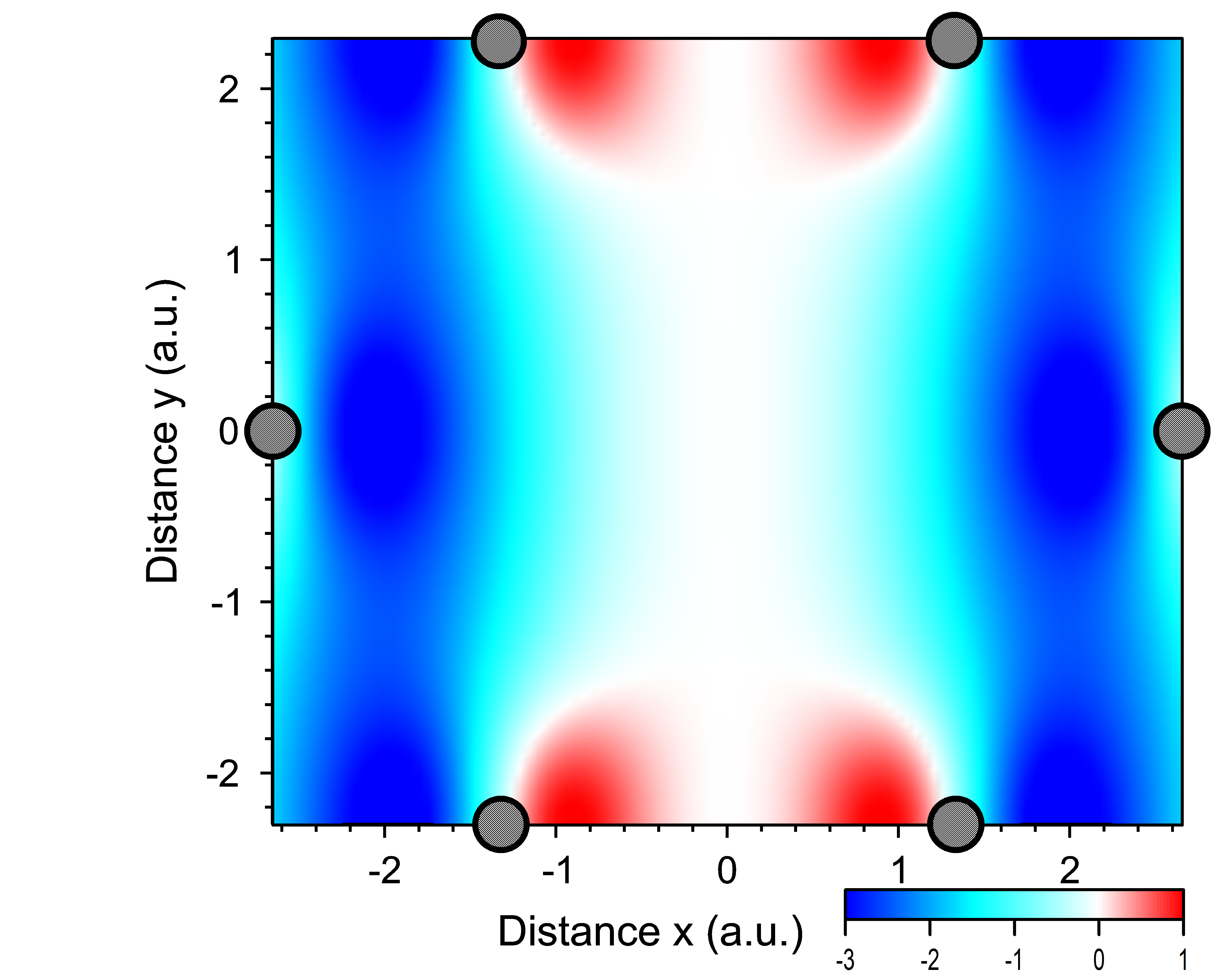}
\end{minipage}
  \hfill
  \begin{minipage}[b]{0.25\textwidth}
\includegraphics[width= \textwidth]{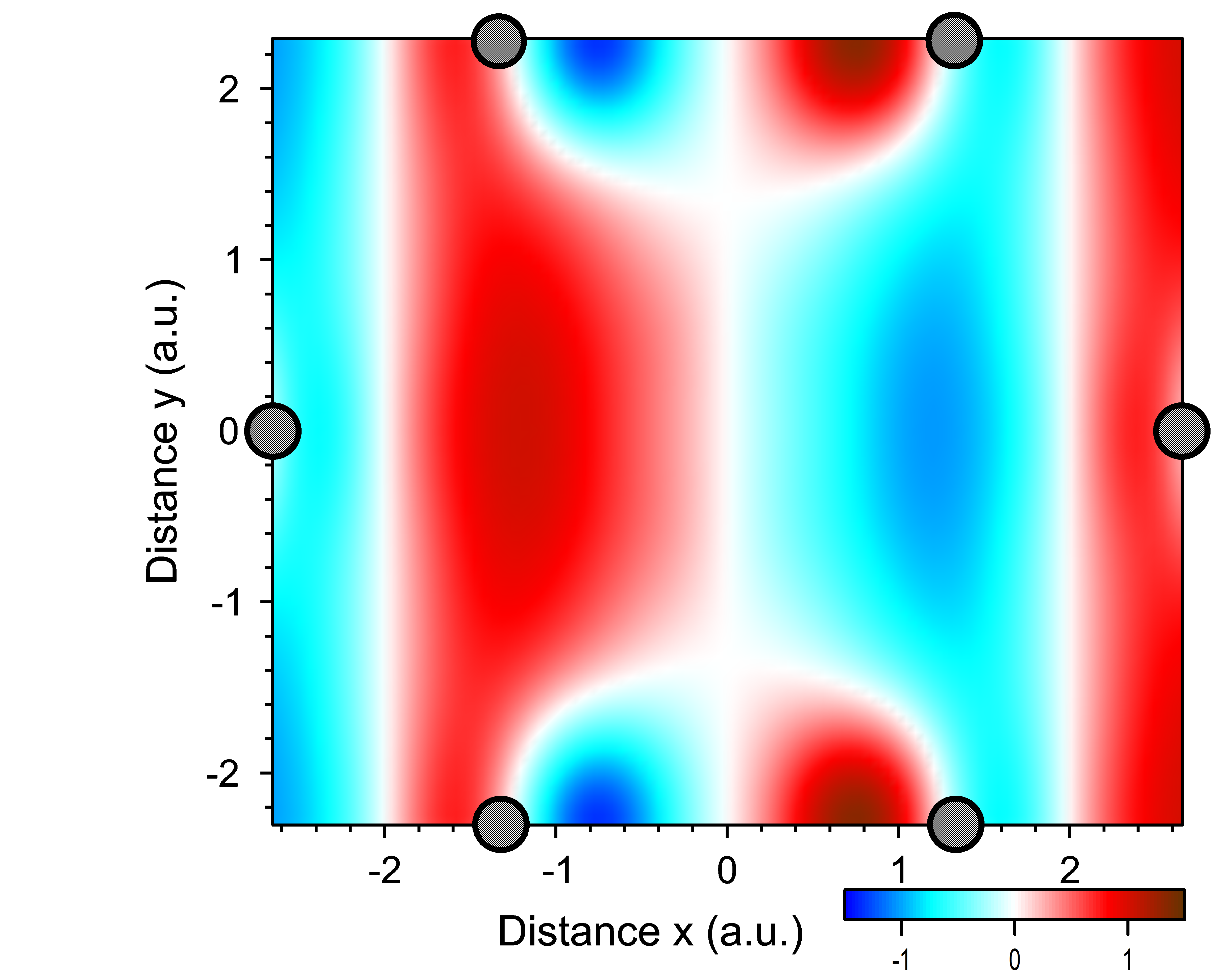}
\end{minipage}
\caption{\label{Psi07}    Real (left) and imaginary (right) parts of the wave function (in arbitrary
units) at the $z = 0$ plane for the seventh band at the $M$ point. }
\end{figure}

\begin{figure}[h]
%\centering
\includegraphics[width= 1\columnwidth]{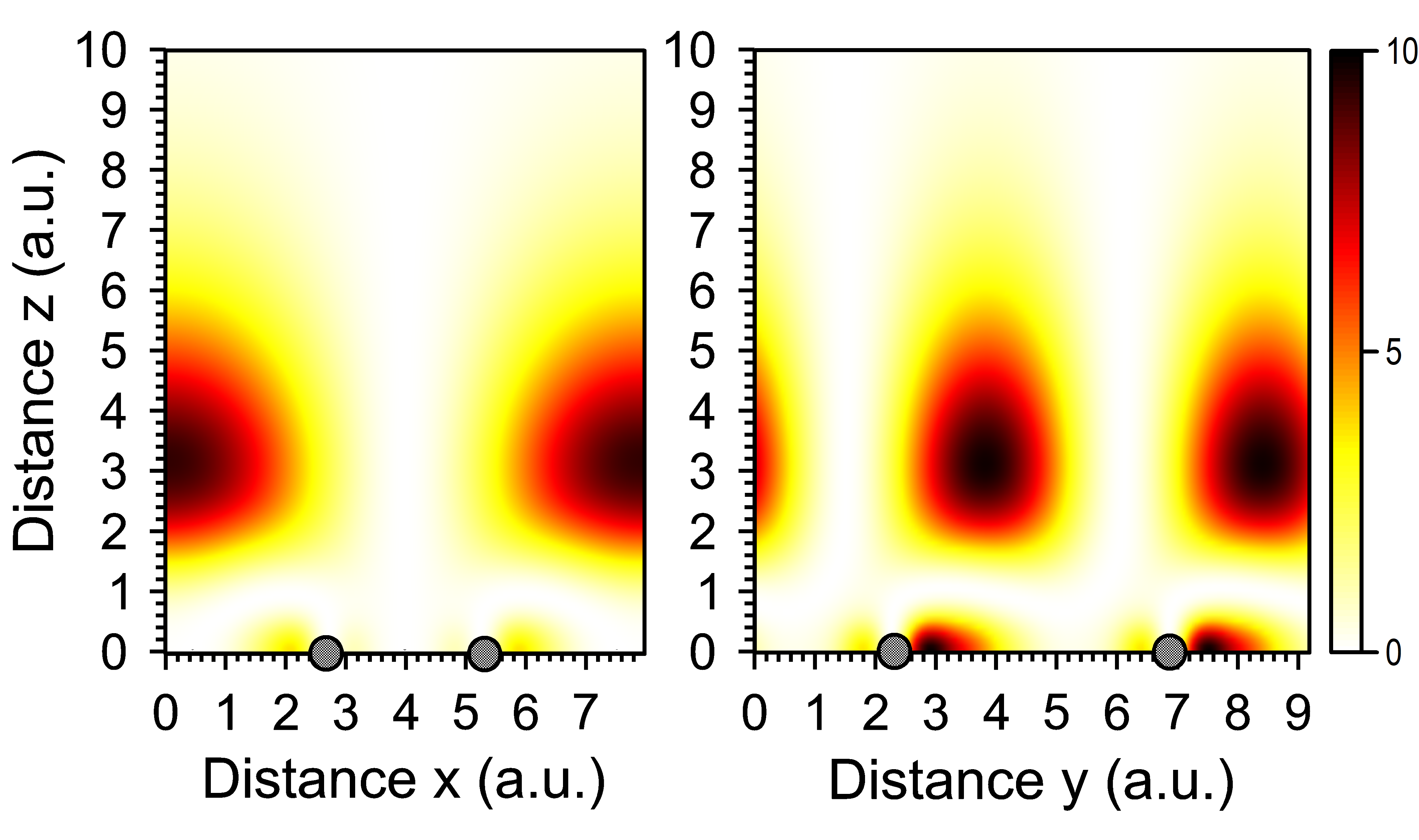}
\caption{\label{Charge08}  Charge-density distribution (in arbitrary units) in $x=0$ plane for the eighth  band at the $M$ point. Filled dots show the carbon ion positions. }
\end{figure}

\begin{figure}[h]
%\centering
\hskip -.5cm
  \begin{minipage}[b]{0.25\textwidth}
  \includegraphics[width= \textwidth]{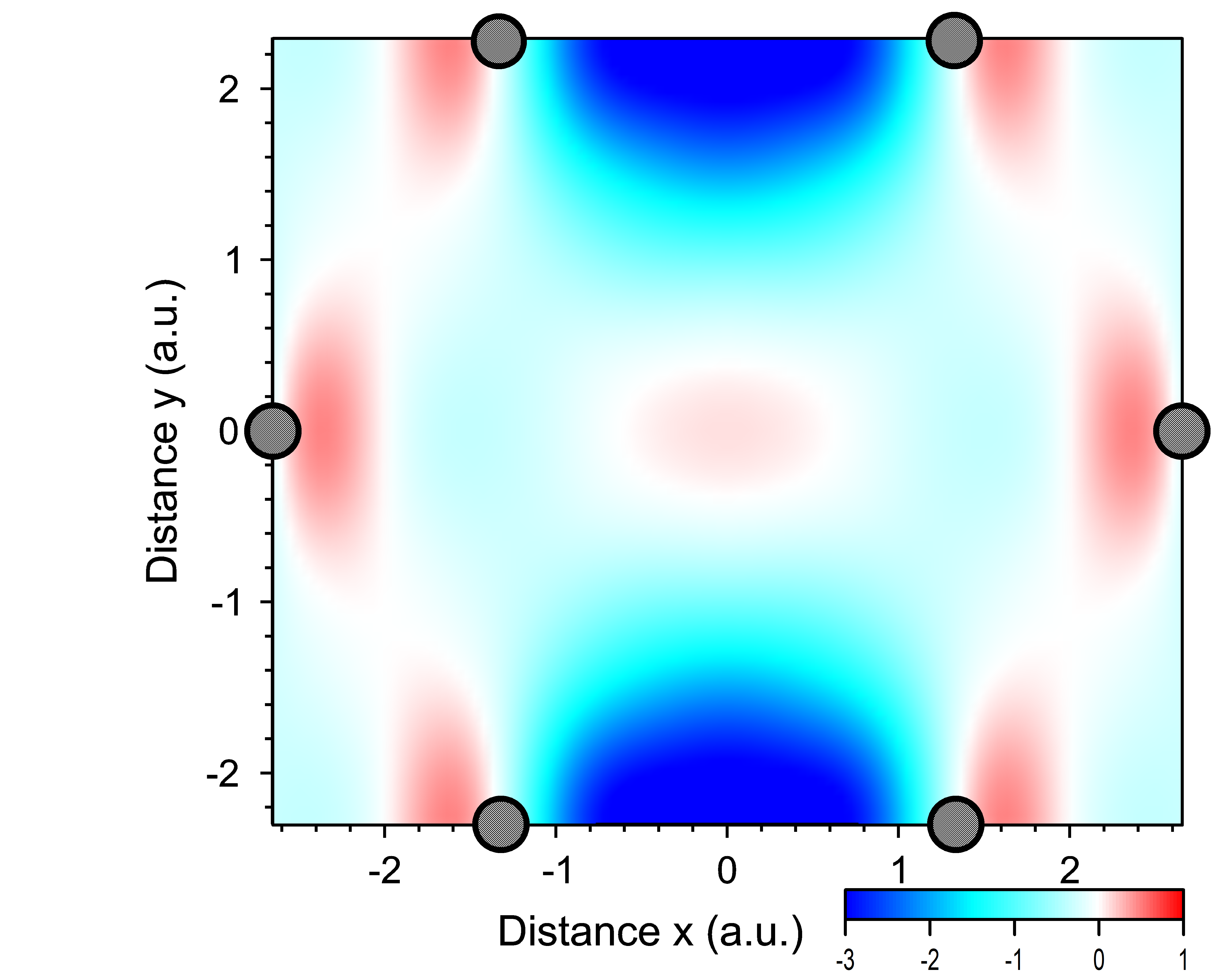}
\end{minipage}
  \hfill
  \begin{minipage}[b]{0.25\textwidth}
\includegraphics[width= \textwidth]{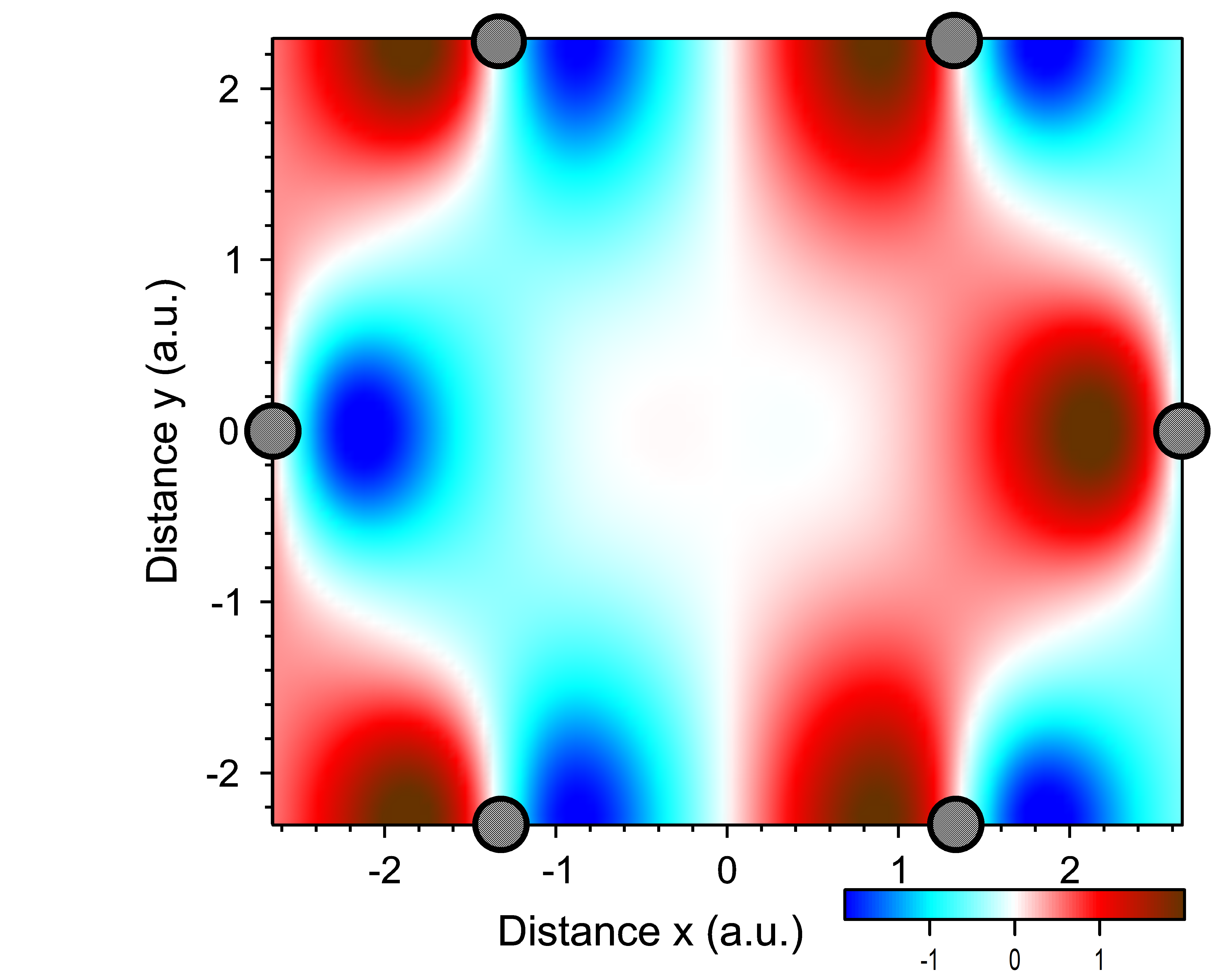}
\end{minipage}
\caption{\label{Psi08}    Real (left) and imaginary (right) parts of the wave function (in arbitrary
units) at the $z = 0$ plane for the eighth band at the $M$ point. }
\end{figure}

The symmetry of each band can be obtained from inspection of the  wave function describing the band (at a given value of wave vector)
obtained as the result of band calculations.
Such analysis was performed in our previous publication for all bands apart from four highest bands at the $M$ point.\cite{sk}
In the present publication we fill this void.
On Figs. \ref{Charge06} -\ref{Psi08} we present the results
of the calculations of the density and wave functions  of the  bands from the sixth to the eighth (counting from below) at the $M$ point.
Inspection of the $z$-dependence of the density shows that these are $\sigma$ bands.
The wave functions of the $\sigma$ bands are plotted at the plane $z=0$. For the  $\pi$ bands,  the wave function is identically equal to zero at the $z=0$ plane, so we plotted the wave function at the plane $z=1$ a.u.

For the seventh band the wave function
 is equal to zero along the $y$ - axis,
which corresponds to the representation $B_{3u}$. (Because the wave function is antisymmetric with respect to reflection, it should be
equal to zero at the axis of reflection.) The wave functions of the sixth and eighth band are different from zero everywhere at the plane,
which is consistent with the representation $A_g$.

\begin{figure}[h]
%\centering
\includegraphics[width= 1\columnwidth]{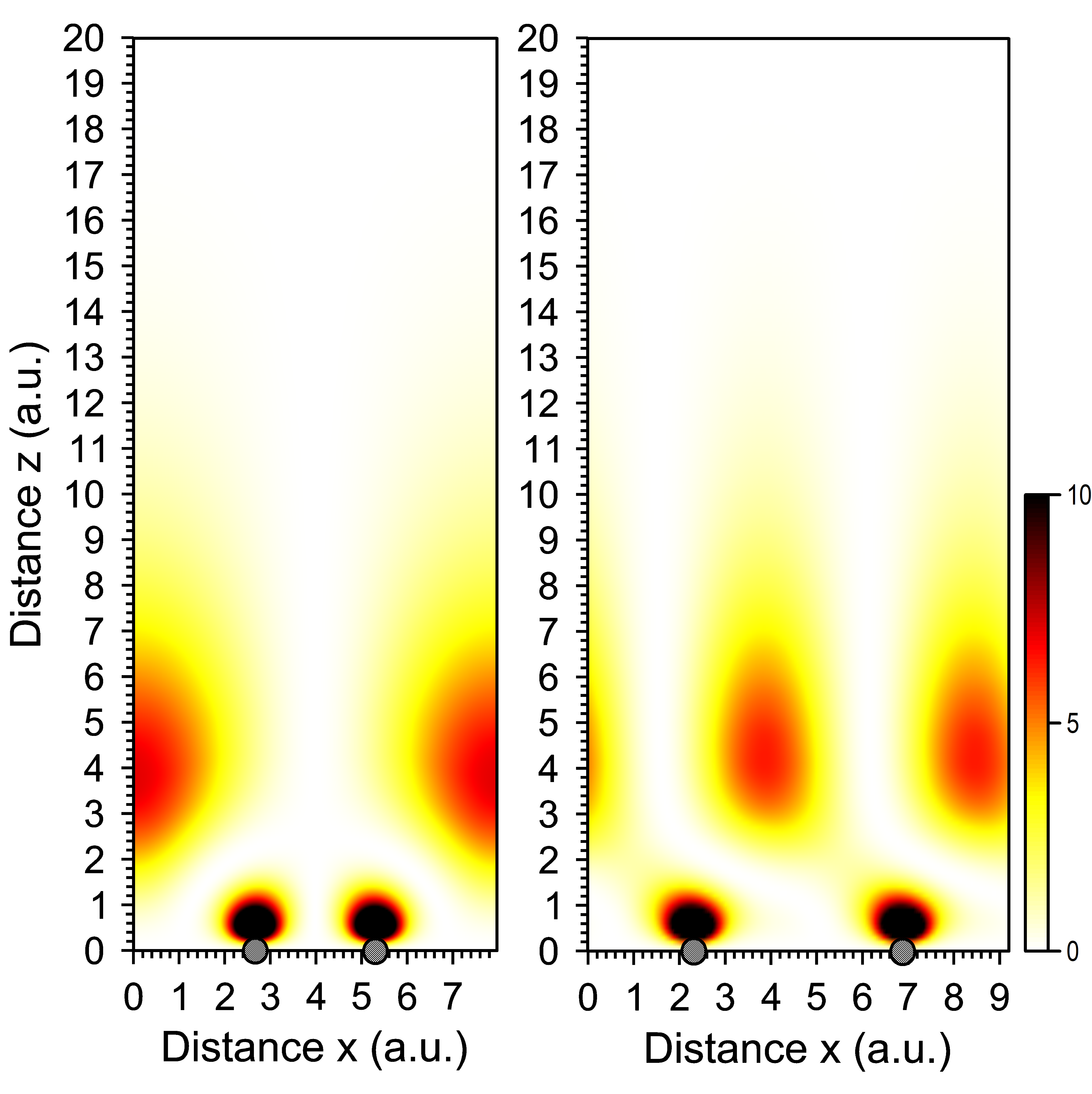}
\caption{\label{Charge09}   Charge-density distribution (in arbitrary units) in $x=0$ plane for the ninth  band at the $M$ point. Filled dots show the carbon ion positions. }
\end{figure}

\begin{figure}[h]
%\centering
\hskip -.5cm
  \begin{minipage}[b]{0.25\textwidth}
  \includegraphics[width= 1\textwidth]{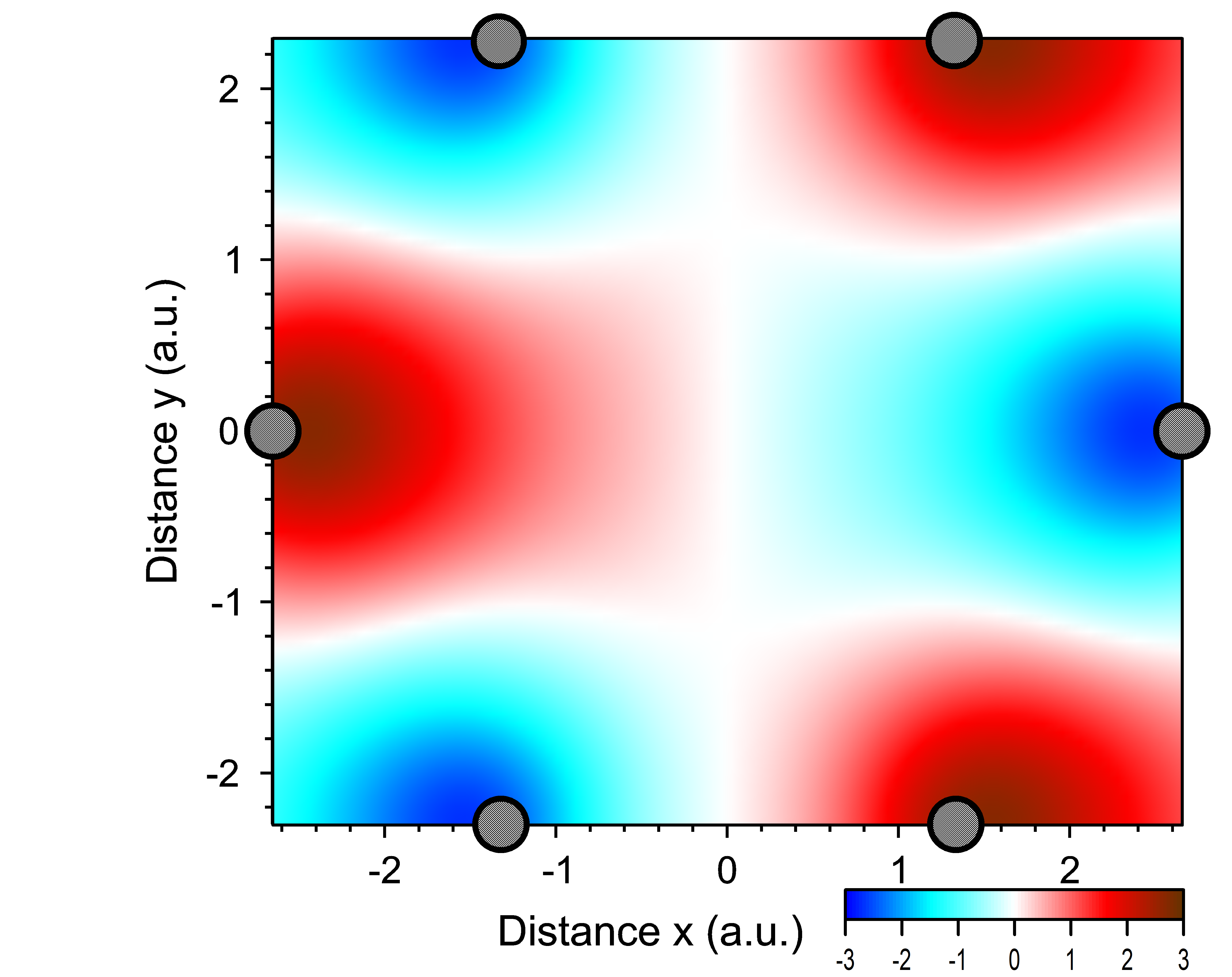}
\end{minipage}
  \hfill
  \begin{minipage}[b]{0.25\textwidth}
\includegraphics[width= 1\textwidth]{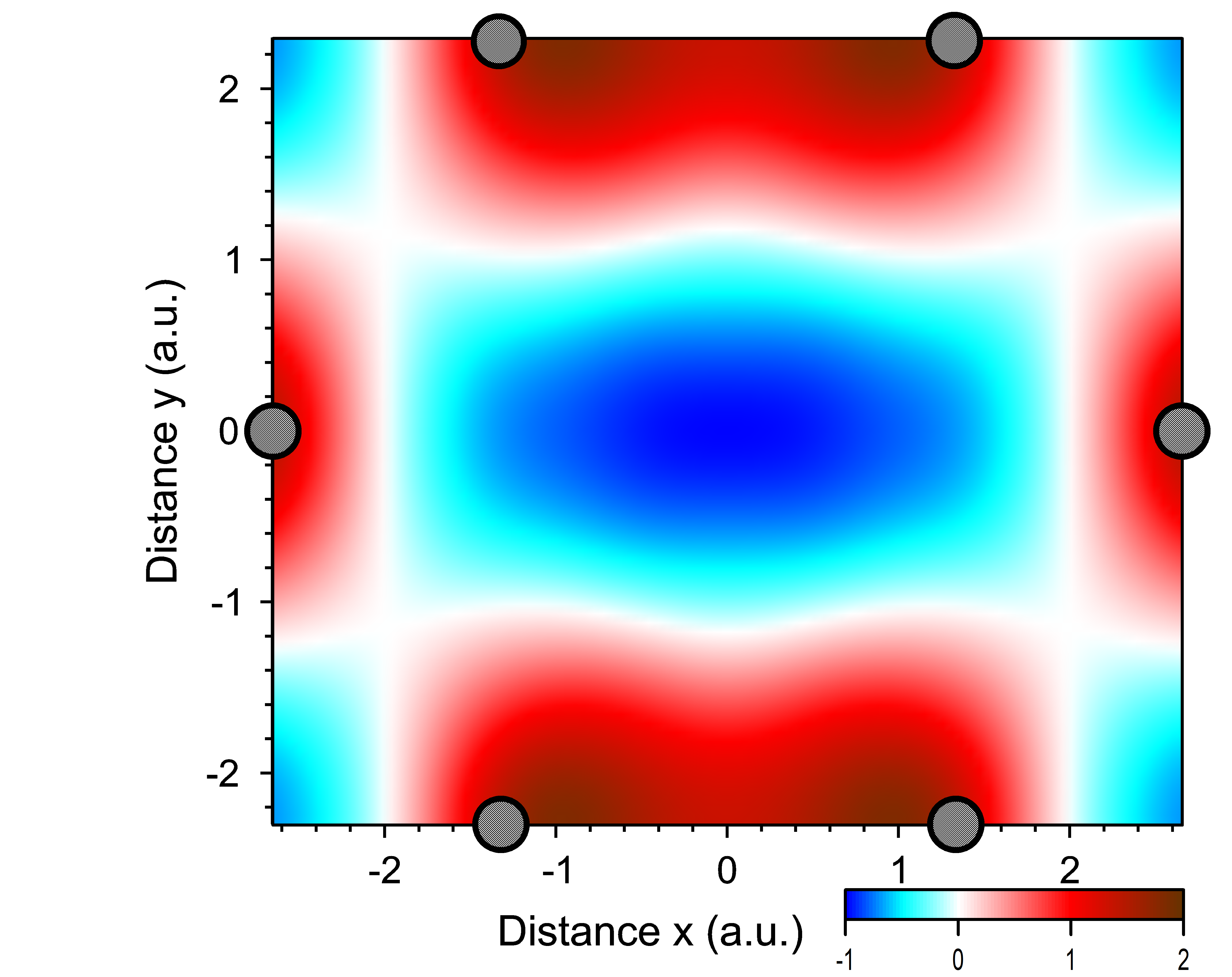}
\end{minipage}
\caption{\label{Psi09}  Real (left) and imaginary (right) parts of the wave function (in arbitrary
units) at the $z = 1$ a.u. plane for the ninth band at the $M$ point. }
\end{figure}

\begin{figure}[h]
%\centering
\includegraphics[width= 1\columnwidth]{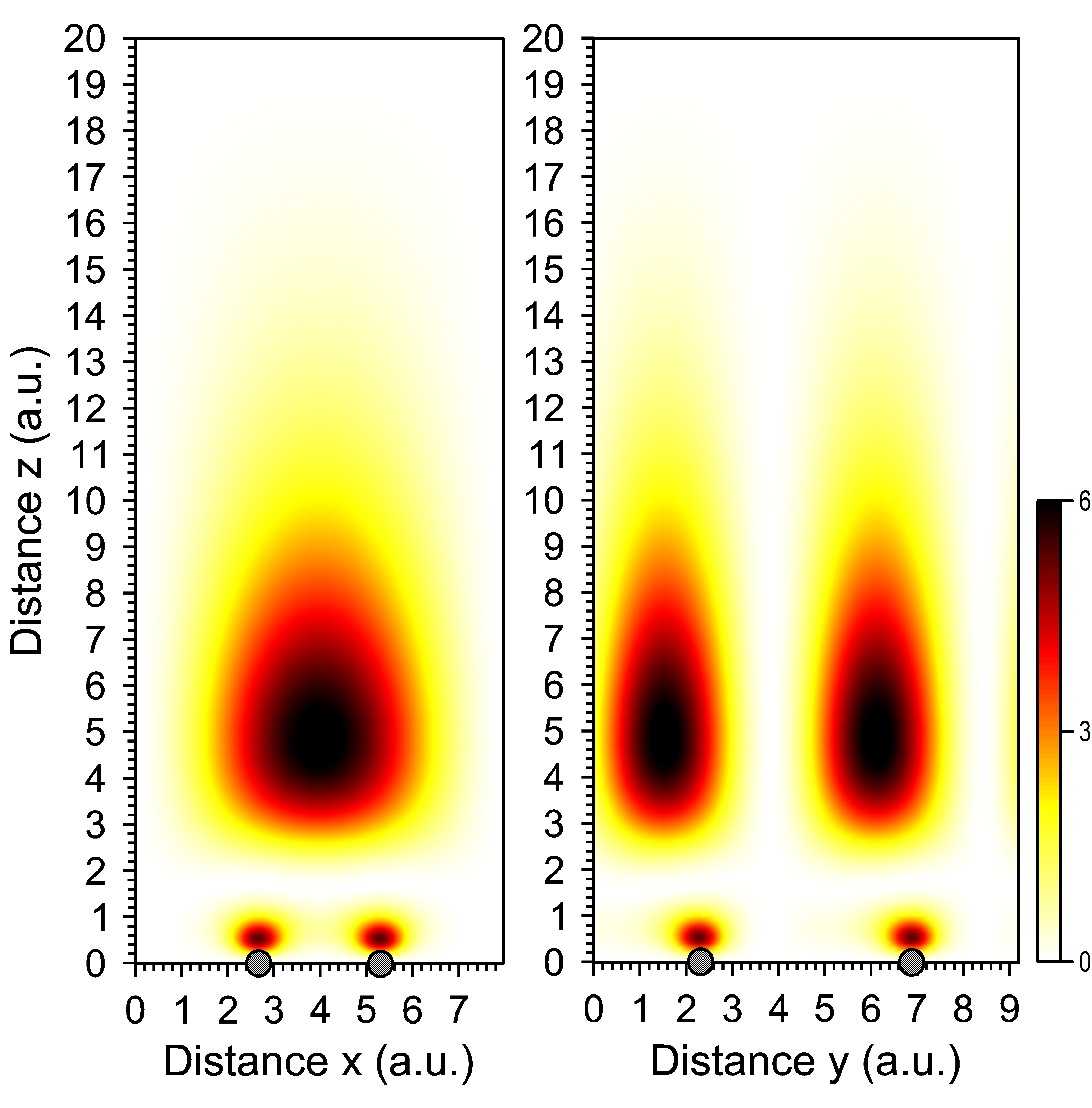}
\caption{\label{Charge10}   Charge-density distribution (in arbitrary units) in $x=0$ plane for the tenth  band at the $M$ point. Filled dots show the carbon ion positions. }
\end{figure}

\begin{figure}[h]
%\centering
\hskip -.5cm
  \begin{minipage}[b]{0.25\textwidth}
  \includegraphics[width= 1\textwidth]{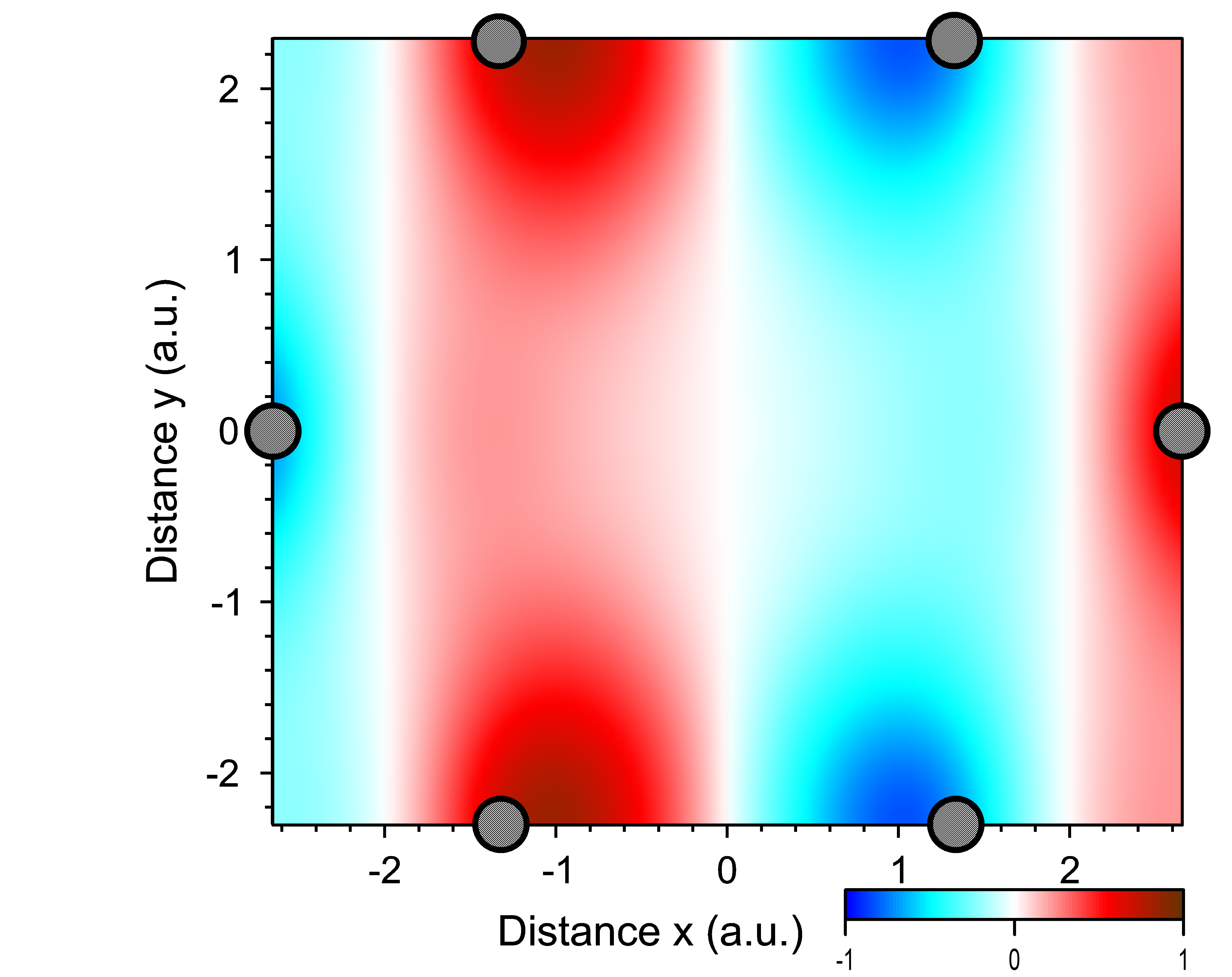}
\end{minipage}
  \hfill
  \begin{minipage}[b]{0.25\textwidth}
\includegraphics[width= 1\textwidth]{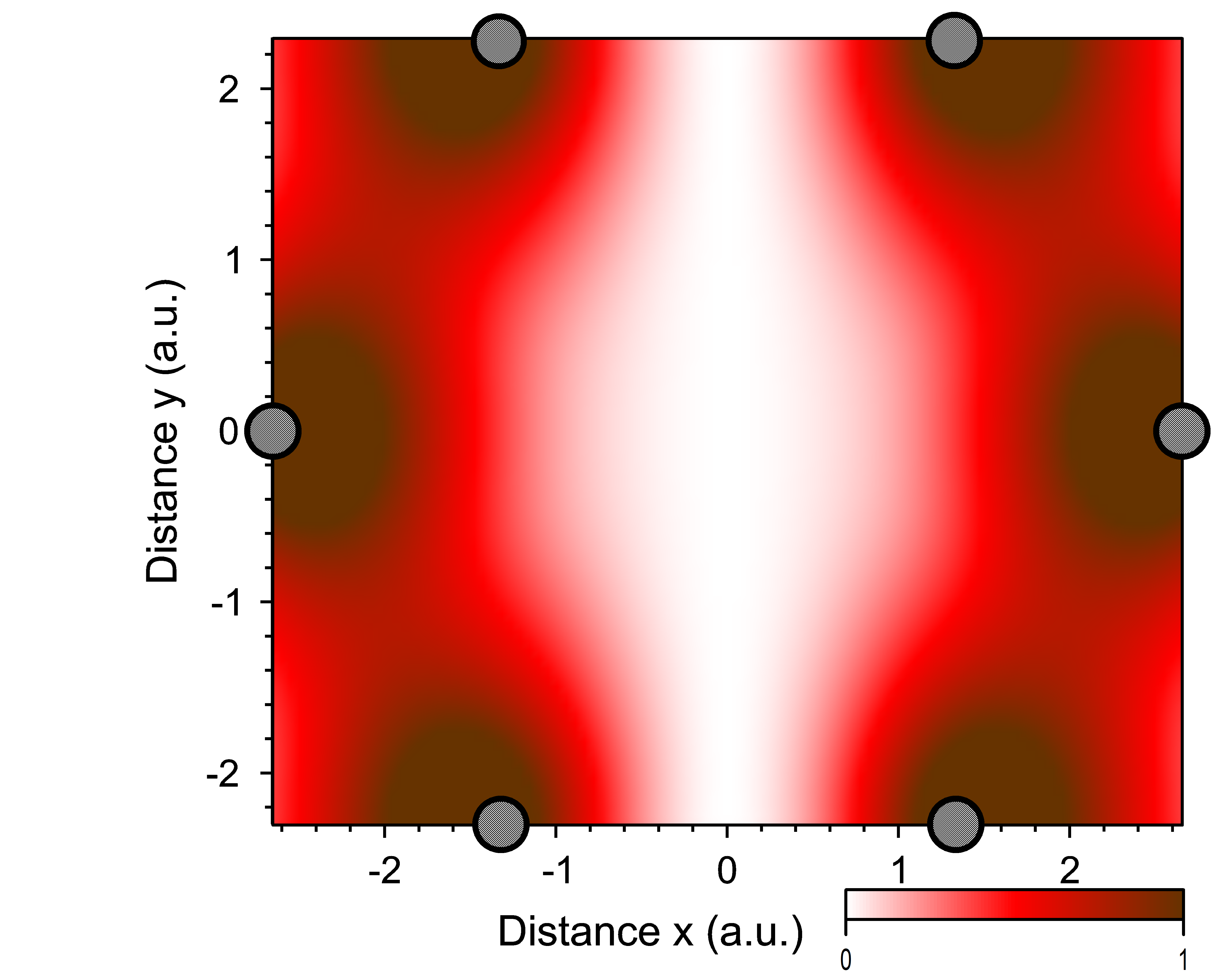}
\end{minipage}
\caption{\label{Psi10}    Real (left) and imaginary (right) parts of the wave function (in arbitrary
units) at the $z = 1$ a.u. plane for the tenth band at the $M$ point. }
\end{figure}

On Figs. \ref{Charge09} and \ref{Charge10} we present the results
of the calculations of the density   of the   ninth and tenth bands.
The wave function is identically equal to zero at the $z=0$ plane,
so it is $\pi$ band. To find the symmetry of the bands on Figs. \ref{Psi09} and \ref{Psi10} we plot the wave function at the plane $z=1$ a.u.
 The wave function of the ninth band is different from zero everywhere at the plane,
which is consistent with the representation $B_{1u}$. For the tenth band the wave function
 is equal to zero along the $y$ - axis,
which corresponds to the representation $B_{3g}$.

The calculated symmetry of the bands can be explained in the framework of both models.
The FEM, in addition, predicts  near degeneracy within the groups of the bands (we call such groups multiplets) and  positions of such groups
relative to each other.

Let us start our analysis from $\sigma$ bands.
At the $\Gamma$ point the TBM with the $|2s,2p_{x,y}>$ basis gives $A_{1g}+E_{2g}+B_{1u}+E_{1u}$ bands.
The  position of the three last bands relative to the three first ones can be understood recalling the distinction between binding and anti-binding orbitals.
Addition to the basis of the $|3s>$
orbital gives additionally $A_{1g}$ and $B_{1u}$ bands. We have to assume that the $B_{1u}$ band is swallowed by the continuum.

The FEM gives at the $\Gamma$ point the lowest band ($A_{1g}$), then the sextuplet $A_{1g}+E_{2g}+B_{1u}+E_{1u}$. The large energy difference is  explained by the fact that the former corresponds to the plane wave with ${\bf k}=0$, and the latter is constructed from plane waves corresponding to the points $\Gamma_1,\dots,\Gamma_6$. The order of the bands in the sextuplet was discussed in Section \ref{ki}. The order within the sextuplet observed on Fig. \ref{fig:bands} can be explained in the framework of the FEM by making plausible assumptions about the lattice (pseudo) potential (see Section \ref{ki}).

At the $M$ point the TBM  gives three times $A_g+B_{3u}$ bands, constructed from $|2s>$, $|2p_x>$ and $|3p_x>$ orbitals respectively, and
 $B_{1g}+B_{2u}$ bands, constructed from $|2p_y>$ orbitals.
We have to assume that $B_{1g}$  and $B_{3u}$ bands are swallowed by the continuum.
The counterintuitive fact is that the band $B_{1g}$ of the $|2p_y>$ origin is swallowed,
while one of the bands of the $|3p_x>$ origin isn't.

The FEM gives at the $M$ point the lowest doublet ($A_g+B_{3u}$), then the higher doublet ($A_{g}+B_{2u}$), and then four still higher bands ($A_g+B_{1g}+B_{2u}+B_{3u}$).
We have to assume that the bands $B_{1g}$ and $B_{2u}$ from the highest quadruplet are  swallowed by the continuum.
The  distances  between the  multiplets  is explained  in the FEM by the fact that the lowest doublet is constructed from the plane waves with the wave vectors corresponding to the points $M$ and $M_2$, the second one - $M_3$ and $M_4$, and the quartet - $M_{5,\dots,8}$.
The weak  the potential $V(x,y)$ should   lead to weak
splitting within each multiplet, and also to  weak splitting of the lowest doublet along the whole $K-M$ line.
The weak splitting of the lowest doublet along the whole $K-M$ line (including the $M$ point), and the weak
splitting of the highest doublet at the $M$ point is what we see on Fig. \ref{fig:bands}.
To be honest we must notice that strong splitting of the second doublet at the $M$ point doesn't agree well with the idea of weak (pseudo) potential $V(x,y)$.

At the $K$ point the TBM gives $A_{1}'+A_{2}$ bands of $|2s>$ origin, the bands with the same symmetry
of $|3s>$ origin, and twice $E'$ bands of of $|p_{x,y}>$ origin.
To be in line with the band calculations we have to assume that one of the $|2s>$ bands is swallowed by the continuum,
while one of the $|3s>$ bands isn't.

The FEM gives  at the $K$ point two triplets  with identical symmetry ($A_1'+E'$).
Large distance between the triplets
is  explained by the fact that the fist triplet is constructed from the plane waves corresponding to the points $K,K_2,K_3$, and the second
triplet is constructed from the plane waves corresponding to the points $K_4,K_5,K_6$.
The assumption of weak $V(x,y)$ potential leads to prediction that the bands within each triplet will be
only weakly split.

The FEM predicts relative the positions of the bands at the line $K-M$: close doublet,  higher a single band, still higher the next single band, and then still higher another close doublet. This prediction corresponds to what we see on Fig. \ref{fig:bands}.

Now let us come to $\pi$ bands.
At the $\Gamma$ point the TBM  gives $A_{2u}+B_{2g}$ bands constructed from $|2p_z>$ orbitals and bands with the same symmetry constructed from $|3p_z>$ orbitals. We have to assume that $B_{2g}$  $|3p_z>$ bands is swallowed by the continuum.
The FEM gives $A_{2u}$ band and sextuplet $A_{2u}+B_{2g}+E_{1u}+E_{2g}$. We have to assume that $E_{1u}$ and
 $E_{2g}$ bands are swallowed by the continuum.

At the $M$ point the TBM  gives $B_{1u}+B_{2g}$ bands, constructed from $|2p_z>$ orbitals, and bands with the same symmetry constructed from $|3p_z>$ orbitals. We have to assume that $B_{2g}$  $|3p_z>$ band is swallowed by the continuum.
The FEM gives lower doublet $B_{1u}+B_{2g}$ and the second doublet  $B_{1u}+B_{3g}$.
%We have to assume that $B_{3g}$ band is swallowed by the continuum.

The $K$ point is especially problematic to TBM. More specifically, the two bands merging at the Fermi level and realizing representation $E''$ are well described as constructed from the $|2p_z>$ orbitals. The problem is with the higher $\pi$ band. Like it was shown in Section \ref{ti},
to describe the nondegenerate $\pi$ band at the $K$ point in the framework of the TBM we need
$\left|3d_{xz}\right>,\left|3d_{yz}\right>$ orbitals.
But this choice leaves unanswered the question: Why,  $\pi$ band
constructed from $|3d>$ orbitals turns out to be lower than that
constructed from $|3p_z>$ orbitals? Probably it can be explained by its hybridization with the scattering resonances predicted\cite{nakrprb13} and observed\cite{jokanc15,wiloprb16,krmaapl17} recently in graphene. It would be interesting to clarify this point in the future.

On the other hand, the FEM predicts at the $K$ point the triplet $E''+A_2''$ which we clearly see on Fig. \ref{fig:bands}.
To be honest we must notice that strong splitting of between the $A_2''$ and $E''$ bands doesn't agree well with the idea of weak (pseudo) potential $V(x,y)$.

Looking at the  bands at the line $K-M$ on Fig. \ref{fig:bands} one sees similarity between $\pi$ bands and four lowest $\sigma$ ones.
The higher $\pi$ bands are swallowed by the continuum.
Comparing  the two alternative approaches to the symmetry classification of the electron bands, we must say that their predictions are complimentary.

\section{Conclusions}

We presented the symmetry labelling of all electron bands  in graphene obtained by combining
numerical band calculations and analytical analysis based on group theory.
The emphasize was on the comparison of the predictions of the  tight-binding
and (nearly) free electron models. The predictions of these two models were found to be complimentary to each others and agreeing well with the results of numerical band calculations.

\section*{Acknowledgments}

The work on this paper started  during E.K. visit to
Max-Planck-Institut f\"{u}r Physik komplexer Systeme in December of 2019 and January of 2020.
E.K. cordially thanks the Institute for the hospitality extended to him during
that and all his previous visits.

V.M.S. acknowledges support from the Project of the Basque Government for consolidated groups of the Basque University, through the Department of Universities (Q-NANOFOT IT1164-19) and from the Spanish Ministry of Science and Innovation (Grant No. PID2019--105488GB--I00).

We are grateful to G. J. Ferreira, G. G. Naumis and R.-J. Slager for  bringing our attention to their papers.

\end{document}